\documentclass[groupedaddress,aps,pra,superscriptaddress,showpacs,twocolumn,prl]{revtex4}%
\usepackage{epsfig,dsfont,amssymb,amsmath,amsthm,amsfonts,amsbsy,mathrsfs}
\usepackage{graphicx, color}
\usepackage{epstopdf}
\usepackage{mathdots}
\usepackage{float}
\begin{document}

\title{Finer Distribution of Quantum Correlations among Multiqubit Systems}

\author{Zhi-Xiang Jin}
\email{jzxjinzhixiang@126.com}
\affiliation{School of Mathematical Sciences,  Capital Normal University,  Beijing 100048,  China}
\affiliation{School of Physics, University of Chinese Academy of Sciences, Yuquan Road 19A, Beijing 100049, China}
\author{Shao-Ming Fei}
\email{feishm@cnu.edu.cn}
\affiliation{School of Mathematical Sciences,  Capital Normal University,  Beijing 100048,  China}
\affiliation{Max-Planck-Institute for Mathematics in the Sciences, Leipzig 04103, Germany}

\begin{abstract}
We study the distribution of quantum correlations characterized by
monogamy relations in multipartite systems. By using the Hamming weight of the binary vectors associated with the subsystems,
we establish a class of monogamy inequalities for multiqubit entanglement based on the $\alpha$th ($\alpha\geq 2$) power of concurrence,
and a class of polygamy inequalities for multiqubit entanglement in terms of the $\beta$th ($0\leq \beta\leq2$) power of concurrence and concurrence of assistance.
Moveover, we give the monogamy and polygamy inequalities for general quantum correlations.
Application of these results to quantum correlations like squared convex-roof extended negativity (SCREN),
entanglement of formation and Tsallis-$q$ entanglement gives rise to either
tighter inequalities than the existing ones for some classes of quantum states
or less restrictions on the quantum states. Detailed examples are presented.
\end{abstract}
\maketitle

\section{INTRODUCTION}
Due to the essential roles played in quantum communication and quantum information processing, quantum entanglement \cite{MAN,RPMK,FMA,KSS,HPB,HPBO,JIV,CYSG} has been the subject of many recent studies in recent years. The study of quantum entanglement from various viewpoints has been a very active area and has led to many impressive results.
As the key resources, quantum entanglement has been used many quantum communication protocols such as superdense coding \cite{bs}, quantum teleportation \cite{bbc}, quantum cryptography  \cite{MP}, remote-state preparation \cite{pak}, and quantum computational tasks such as the one-way quantum computer \cite{rh}.
As one of the fundamental differences between quantum and classical correlations, an essential property of entanglement is that a quantum system entangled with one of other subsystems limits its entanglement with the remaining ones. The monogamy relations give rise to the distribution of entanglement in the multipartite quantum systems. Moreover, the monogamy property has emerged as the ingredient in the security analysis of quantum key distribution \cite{MP}.	

The monogamy inequalities are important relations satisfied by the multipartite quantum entanglement. In the classical scenario, the fact that two systems sharing some correlations does not prevent them from being correlated with a third party. Nevertheless, two maximally entangled quantum systems cannot be entangled with a third one. Generally, the more the entanglement between two systems, the less the entanglement with the rest systems.
For a tripartite system $A$, $B$ and $C$, the usual monogamy of an entanglement measure $\mathcal{E}$ implies that \cite{MK} the entanglement between $A$ and $BC$ satisfies $\mathcal{E}_{A|BC}\geq \mathcal{E}_{AB} +\mathcal{E}_{AC}$. However, such monogamy relations are not always satisfied by all entanglement measures for all quantum states. In fact, it has been shown that the squared concurrence $C^2$ \cite{TJ,YKM} and entanglement of formation $E^2$ \cite{TR} satisfy the monogamy relations for multi-qubit states. The monogamy inequality was further generalized to various entanglement measures such as continuous-variable entanglement \cite{agf,hai,agi}, squashed entanglement \cite{MK,cwa,yd}, entanglement negativity \cite{ofh,kds,hvg,ckj,lly}, Tsallis-q entanglement \cite{kjs,kjsg}, and Renyi-entanglement \cite{ksb,cdm,wmv}.
Monogamy relations characterize the distributions of quantum correlations in multipartite systems and play a crucial role in the security of quantum cryptography. Tighter monogamy relations imply finer characterizations of the quantum correlation distributions, which tightens the security bounds in quantum cryptography \cite{ggt}.

In this paper, we provide a finer characterization of multiqubit entanglement in terms of concurrence.
By using the Hamming weight of the binary vectors related to the subsystems, we establish a class of monogamy inequalities for
multiqubit entanglement based on the $\alpha$th power of concurrence for $\alpha\geq 2$. For $0\leq \beta\leq2$, we establish
a class of polygamy inequalities for multiqubit entanglement in terms of the $\beta$th power of concurrence and concurrence of assistance.
We further show that our class of monogamy and polygamy inequalities hold in a tighter way than those provided before.
Then we give the monogamy and polygamy inequalities for general quantum correlations, which can applied to SCREN, entanglement of formation and Tsallis-$q$ entanglement, and give rise to
tighter inequalities than the existing ones for some classes of quantum states,
or to monogamy relations with less constraints on quantum states. Moreover, we take SCREN as an example and show the advantage of the general monogamy and polygamy inequalities.
We also show that our monogamy inequalities still valid for the counterexamples of the tangle-based monogamy inequality, where at least one local dimension is larger than two.

\section{MONOGAMY RELATIONS FOR CONCURRENCE BASED ON Hamming weight}

We first consider the monogamy inequalities satisfied by the concurrence. Let $\mathds{H}_X$ denote a discrete finite-dimensional complex vector space associated with a quantum subsystem $X$.
For a bipartite pure state $|\psi\rangle_{AB}\in\mathds{H}_A\otimes \mathds{H}_B$, the concurrence is given by \cite{AU,PR,SA}, $C(|\psi\rangle_{AB})=\sqrt{{2\left[1-\mathrm{Tr}(\rho_A^2)\right]}}$,
where $\rho_A$ is the reduced density matrix obtained by tracing over the subsystem $B$, $\rho_A=\mathrm{Tr}_B(|\psi\rangle_{AB}\langle\psi|)$. The concurrence for a bipartite mixed state $\rho_{AB}$ is defined by the convex roof extension,
$C(\rho_{AB})=\min_{\{p_i,|\psi_i\rangle\}}\sum_ip_iC(|\psi_i\rangle)$,
where the minimum is taken over all possible decompositions of $\rho_{AB}=\sum\limits_{i}p_i|\psi_i\rangle\langle\psi_i|$, with $p_i\geq0$ and $\sum\limits_{i}p_i=1$ and $|\psi_i\rangle\in \mathds{H}_A\otimes \mathds{H}_B$.

For a tripartite state $|\psi\rangle_{ABC}$, the concurrence of assistance is defined by \cite{TFS, YCS},
\begin{eqnarray*}
C_a(|\psi\rangle_{ABC})\equiv C_a(\rho_{AB})=\max\limits_{\{p_i,|\psi_i\rangle\}}\sum_ip_iC(|\psi_i\rangle),
\end{eqnarray*}
where the maximum is taken over all possible decompositions of $\rho_{AB}=\mathrm{Tr}_C(|\psi\rangle_{ABC}\langle\psi|)=\sum\limits_{i}p_i|\psi_i\rangle_{AB}\langle\psi_i|.$ When $\rho_{AB}=|\psi\rangle_{AB}\langle\psi|$ is a pure state, one has $C(|\psi\rangle_{AB})=C_a(\rho_{AB})$.

For an $N$-qubit state $\rho_{AB_1\cdots B_{N-1}}\in \mathds{H}_A\otimes \mathds{H}_{B_1}\otimes\cdots\otimes \mathds{H}_{B_{N-1}}$, the concurrence $C(\rho_{A|B_1\cdots B_{N-1}})$ of the state $\rho_{A|B_1\cdots B_{N-1}}$, viewed as a bipartite state under the partition $A$ and $B_1,B_2,\cdots, B_{N-1}$, satisfies the Coffman-Kundu-Wootters (CKW) inequality \cite{TJ,YKM},
\begin{eqnarray}\label{C2}
  C^2(\rho_{A|B_1,B_2\cdots,B_{N-1}})\geq \sum_{i=1}^{N-1}C^2(\rho_{AB_i}),
\end{eqnarray}
for $\alpha\geq2$, where $\rho_{AB_i}=\mathrm{Tr}_{B_1\cdots B_{i-1}B_{i+1}\cdots B_{N-1}}(\rho_{AB_1\cdots B_{N-1}})$.
The dual inequality in terms of the concurrence of assistance for $N$-qubit states has the form \cite{GSB},
\begin{eqnarray}\label{DCA}
 C^2(\rho_{A|B_1,B_2\cdots,B_{N-1}})\leq \sum_{i=1}^{N-1}{C_a^2}(\rho_{AB_i}).
\end{eqnarray}

It is further improved that for $\alpha\geq2$ \cite{ZXN},
\begin{eqnarray}\label{mo1}
C^{\alpha}(\rho_{A|B_1,B_2\cdots,B_{N-1}})\geq\sum_{i=1}^{N-1}C^\alpha(\rho_{AB_i}).
\end{eqnarray}
Moreover, for $\alpha\geq2$, if $C(\rho_{AB_i})\geq C(\rho_{A|B_{i+1}\cdots B_{N-1}})$ for $i=1, 2, \cdots, N-2$, $N\geq 4$, then \cite{JZX},
\begin{eqnarray}\label{mo2}
&&C^\alpha(\rho_{A|B_1B_2\cdots B_{N-1}})\geq C^\alpha(\rho_{AB_1})\nonumber\\
&&+\frac{\alpha}{2} C^\alpha(\rho_{AB_2})+\cdots+\left(\frac{\alpha}{2}\right)^{N-2}C^\alpha(\rho_{AB_{N-1}})
\end{eqnarray}
and for all $\alpha<0$,
\begin{eqnarray*}\label{l2}
\setlength{\belowdisplayskip}{3pt}
&&C^\alpha(\rho_{A|B_1B_2\cdots B_{N-1}})< \nonumber \\
&&K(C^\alpha(\rho_{AB_1})+C^\alpha(\rho_{AB_2})+\cdots+C^\alpha(\rho_{AB_{N-1}})),
\end{eqnarray*}
where $K=\frac{1}{N-1}$.

Before we present our main results, we first provide some notations and lemmas. For convenience, we denote $C_{AB_i}=C(\rho_{AB_i})$ the concurrence of $\rho_{AB_i}$ and $C_{A|B_1,B_2\cdots,B_{N-1}}=C(\rho_{A|B_1\cdots B_{N-1}})$.

For any non-negative integer $j$ and its binary expansion
\begin{eqnarray*}\label{}
j=\sum_{i=0}^{n-1}j_i2^i,
\end{eqnarray*}
with $\log_2j\leq n$ and $j_i\in \{0,~1\}$, for $i=0,1,\cdots,n-1$, we can always define a unique binary vector $\vec{j}$ associated with $j$,
\begin{eqnarray}\label{wj}
\vec{j}=(j_0,j_1,\cdots,j_{n-1}).
\end{eqnarray}
The Hamming weight $w_H(\vec{j})$ of $\vec{j}$ is defined as the number of $1's$ in $\{j_0,j_1,\cdots,j_{n-1}\}$ \cite{MAN}.

{[\bf Lemma 1]}. \cite {jll} For any real numbers $x$ and $t$, $0\leq t \leq 1$, $x\in [1,\infty)$, we have $(1+t)^x\geq 1+(2^{x}-1)t^x$.

{[\bf Lemma 2]}. \cite {jll} For any $2\otimes2\otimes2^{n-2}$ mixed state $\rho\in \mathds{H}_A\otimes \mathds{H}_{B}\otimes \mathds{H}_{C}$, if $C_{AB}\geq C_{AC}$, we have
\begin{equation}\label{le2}
  C^\alpha_{A|BC}\geq  C^\alpha_{AB}+(2^{\frac{\alpha}{2}}-1)C^\alpha_{AC},
\end{equation}
for all $\alpha\geq2$.

From lemma 2, we have the following theorem, which states that a class of monogamy inequalities of multiqubit entanglement can be established using the $\alpha$-powered concurrence and the Hamming weight of the binary vector related with the
distribution of subsystems.

{[\bf Theorem 1]}. For any $N+1$-qubit state $\rho_{AB_0\cdots B_{N-1}}$ satisfying
\begin{eqnarray}\label{th11}
C_{AB_j}\geq C_{AB_{j+1}}\geq 0,
\end{eqnarray}
for $j=0,1,\cdots N-2$, we have for $\alpha\geq2$,
\begin{eqnarray}\label{th12}
C^\alpha_{A|B_0B_1\cdots B_{N-1}}\geq \sum_{j=0}^{N-1} (2^{\frac{\alpha}{2}}-1)^{w_H(\vec{j})}C^\alpha_{AB_j},
\end{eqnarray}
where $\vec{j}=(j_0,j_1,\cdots,j_{N-1})$ is the vector from the binary representation of $j$ and $w_H(\vec{j})$ is the Hamming weight of $\vec{j}$.

{\sf [Proof].} Without loss of generality, we can always assume that inequality Eq. (\ref{th11}) holds by relabeling the subsystems.
From Eq. (\ref{C2}), it is sufficient to show that
\begin{eqnarray}\label{pfth11}
\left(\sum_{j=0}^{N-1} C^2_{AB_j}\right)^\frac{\alpha}{2}\geq \sum_{j=0}^{N-1} (2^{\frac{\alpha}{2}}-1)^{w_H(\vec{j})}C^\alpha_{AB_j}.
\end{eqnarray}
We first prove the inequality Eq. (\ref{pfth11}) for the case that $N$ is a power of 2, $N = 2^n$, by mathematical induction. For $n = 1$ and a three-qubit state $\rho_{AB_0B_1}$, we have from Lemma 2,
\begin{equation*}\label{}
  C^\alpha_{A|B_0B_1}\geq  C^\alpha_{AB_0}+(2^{\frac{\alpha}{2}}-1)C^\alpha_{AB_1},
\end{equation*}
which is just the inequality (\ref{pfth11}) for this case.

Now assume that inequality (\ref{pfth11}) is true for $N=2^{n-1}$, $n \geq 2$. Then for an $(N+1)$-qubit state $\rho_{AB_0\cdots B_{N-1}}$, we have
\begin{eqnarray}\label{pfth12}
&&\left(\sum_{j=0}^{N-1} C^2_{AB_j}\right)^\frac{\alpha}{2}\nonumber\\
&&= \left(\sum_{j=0}^{2^{n-1}-1} C^2_{AB_j}+\sum_{j=2^{n-1}}^{2^n-1} C^2_{AB_j}\right)^\frac{\alpha}{2}\nonumber\\
&&= \left(\sum_{j=0}^{2^{n-1}-1} C^2_{AB_j}\right)^\frac{\alpha}{2}\left(1+\frac{\sum_{j=2^{n-1}}^{2^n-1} C^2_{AB_j}}{\sum_{j=0}^{2^{n-1}-1} C^2_{AB_j}}\right)^\frac{\alpha}{2}.
\end{eqnarray}

Due to inequality (\ref{th11}), we have
\begin{eqnarray}\label{pfth13}
\sum_{j=2^{n-1}}^{2^n-1} C^2_{AB_j}\leq\sum_{j=0}^{2^{n-1}-1} C^2_{AB_j}.
\end{eqnarray}
By using Lemma 1 we have
\begin{eqnarray}\label{pfth14}
&&\left(\sum_{j=0}^{N-1} C^2_{AB_j}\right)^\frac{\alpha}{2}\nonumber\\
&& \geq\left(\sum_{j=0}^{2^{n-1}-1} C^2_{AB_j}\right)^\frac{\alpha}{2}+(2^{\frac{\alpha}{2}}-1)\left(\sum_{j=2^{n-1}}^{2^n-1} C^2_{AB_j}\right)^\frac{\alpha}{2}.
\end{eqnarray}
Here, the induction hypothesis that (\ref{pfth11}) is true for $N=2^{n-1}$ implies that
\begin{eqnarray}\label{pfth15}
\left(\sum_{j=0}^{2^{n-1}-1} C^2_{AB_j}\right)^\frac{\alpha}{2}\geq \sum_{j=0}^{2^{n-1}-1} (2^{\frac{\alpha}{2}}-1)^{w_H(\vec{j})}C^\alpha_{AB_j}.
\end{eqnarray}
Note that the last summation in inequality (\ref{pfth14}) is also a summation of $2^{n-1}$ terms
from $j=2^{n-1}$ to $2^n-1$. From (\ref{pfth15}) we obtain
\begin{eqnarray}\label{pfth16}
\left(\sum_{j=2^{n-1}}^{2^{n}-1} C^2_{AB_j}\right)^\frac{\alpha}{2}\geq \sum_{j=2^{n-1}}^{2^{n}-1} (2^{\frac{\alpha}{2}}-1)^{w_H(\vec{j})-1}C^\alpha_{AB_j}.
\end{eqnarray}
From inequality (\ref{pfth14}), (\ref{pfth15}) and (\ref{pfth16}),
we have
\begin{eqnarray}\label{pfth17}
\left(\sum_{j=0}^{2^{n}-1} C^2_{AB_j}\right)^\frac{\alpha}{2}\geq \sum_{j=0}^{2^{n}-1} (2^{\frac{\alpha}{2}}-1)^{w_H(\vec{j})}C^\alpha_{AB_j} ,
\end{eqnarray}
which proves the inequality (\ref{pfth11}).

Now we consider the case of arbitrary positive integer $N$. One can always assume that
$0\leq N\leq 2^n$ for some $n$. Let us consider the following $(2^n+1)$-qubit state,
\begin{eqnarray}\label{pfth18}
\rho'_{AB_0\cdots B_{2^n-1}}=\rho_{AB_0\cdots B_{N-1}}\otimes \delta_{B_N\cdots B_{2^n-1}},
\end{eqnarray}
where $\delta_{B_N\cdots B_{2^n-1}}$ is an arbitrary $(2^n -N )$-qubit state.

As $\rho'_{AB_0\cdots B_{2^n-1}}$ is a $(2^n +1)$-qubit state, inequality (\ref{pfth17}) leads to
\begin{eqnarray}\label{pfth19}
C^\alpha(\rho'_{A|B_0B_1\cdots B_{2^n-1}})\geq \sum_{j=0}^{2^n-1} (2^{\frac{\alpha}{2}}-1)^{w_H(\vec{j})}C^\alpha(\sigma_{AB_j}),
\end{eqnarray}
where $\sigma_{AB_j}$ is the two-qubit reduced density matrix of $\rho'_{AB_0\cdots B_{2^n-1}}$, $j=0,1,\cdots,2^n-1$. Taking into account the following obvious facts:
$C(\rho'_{A|B_0B_1\cdots B_{2^n-1}})=C(\rho_{A|B_0B_1\cdots B_{N-1}})$,
$C(\sigma_{AB_j})=0$ for $j=N,\cdots,2^n-1$, and $\sigma_{AB_j}=\rho_{AB_j}$
for $j=0,1,\cdots,N-1$, we have
\begin{eqnarray}\label{pfth113}
&&C^\alpha(\rho_{A|B_0B_1\cdots B_{N-1}})\nonumber\\
&&=C^\alpha(\rho'_{A|B_0B_1\cdots B_{2^n-1}})\nonumber\\
&&\geq \sum_{j=0}^{2^n-1} (2^{\frac{\alpha}{2}}-1)^{w_H(\vec{j})}C^\alpha(\sigma_{AB_j})\nonumber\\
&&=\sum_{j=0}^{N-1} (2^{\frac{\alpha}{2}}-1)^{w_H(\vec{j})}C^\alpha(\rho_{AB_j}),
\end{eqnarray}
which completes the proof. $\Box$

{\sf [Remark 1].} We establish new monogamy relations in terms of the Hamming weight for arbitrary $N+1$-qubit states. Since $(2^{\frac{\alpha}{2}}-1)^{w_H(\vec{j})}\geq 1$ for any $\alpha\geq 2$, we have
\begin{eqnarray}\label{pfth114}
C^\alpha_{A|B_0B_1\cdots B_{N-1}}&&\geq\sum_{j=0}^{N-1} (2^{\frac{\alpha}{2}}-1)^{w_H(\vec{j})}C^\alpha_{AB_j}\nonumber\\
&&\geq\sum_{j=0}^{N-1} C^\alpha_{AB_j}.
\end{eqnarray}
Therefore, inequality (\ref{th12}) of Theorem 1 is generally tighter than the
inequality (\ref{mo1}). Compared with the inequality (\ref{mo2}), which is only valid for some special states satisfying the conditions $C_{AB_i}\geq C_{A|B_{i+1}\cdots B_{N-1}}$, $i=1,\cdots,N-2$,
our inequality (\ref{th12}) is satisfied for any quantum states.

In fact, the inequality (\ref{th12}) can be further improved to be tighter under some conditions on two-qubit entanglement.

{\bf [Theorem 2]}. For any multiqubit state $\rho_{AB_0\cdots B_{N-1}}$ such that
\begin{eqnarray}\label{th22}
C^2_{AB_i}\geq\sum_{j=i+1}^{N-1} C^2_{AB_j},
\end{eqnarray}
for $i=0,1,\cdots N-2$, we have
\begin{eqnarray}\label{th21}
C^\alpha_{A|B_0B_1\cdots B_{N-1}}\geq\sum_{j=0}^{N-1} (2^{\frac{\alpha}{2}}-1)^jC^\alpha_{AB_j}
\end{eqnarray}
for $\alpha\geq 2$.

{\sf [Proof].}
From the inequality (\ref{le2}), we have
\begin{eqnarray*}\label{pfth21}
&&C^{\alpha}_{A|B_0B_1\cdots B_{N-1}}\nonumber\\
&&\geq  C^{\alpha}_{AB_0}+(2^{\frac{\alpha}{2}}-1) \left(\sum_{j=1}^{N-1}C^2_{AB_j}\right)^\frac{\alpha}{2}\nonumber\\
&&\geq C^{\alpha}_{AB_0}+(2^{\frac{\alpha}{2}}-1)C^{\alpha}_{AB_1}
 +(2^{\frac{\alpha}{2}}-1)^2 \left(\sum_{j=2}^{N-1}C^2_{AB_j}\right)^\frac{\alpha}{2}\nonumber\\
&& \geq \cdots\nonumber\\
&&\geq C^{\alpha}_{AB_0}+(2^{\frac{\alpha}{2}}-1)C^{\alpha}_{AB_1}+\cdots+(2^{\frac{\alpha}{2}}-1)^{N-1}C^{\alpha}_{AB_{N-1}}.
\end{eqnarray*}
$\Box$

For any non-negative integer $j$ and its corresponding binary vector $\vec{j}$ in Eq. (\ref{wj}),
the Hamming weight $w_H(\vec{j})$ is bounded above by $\log_2j$. Thus, we have
\begin{eqnarray}\label{pfth22}
w_H(\vec{j})\leq \log_2j\leq j,
\end{eqnarray}
which gives rise to
\begin{eqnarray}\label{pfth23}
&&C^\alpha_{A|B_0B_1\cdots B_{N-1}}\nonumber\\
&&\geq\sum_{j=0}^{N-1} (2^{\frac{\alpha}{2}}-1)^jC^\alpha_{AB_j}\nonumber\\
&&\geq\sum_{j=0}^{N-1} (2^{\frac{\alpha}{2}}-1)^{w_H(\vec{j})}C^\alpha_{AB_j},
\end{eqnarray}
for any $\alpha\geq 2$. In other words, inequality (\ref{th21}) of Theorem 2 is tighter than the inequality (\ref{th11}) of Theorem 1 for any $\alpha\geq 2$ and any multiqubit state $\rho_{A|B_0B_1\cdots B_{N-1}}$ satisfying the condition (\ref{th22}).

Recently, another class of multiqubit monogamy inequalities in terms of the $\alpha$th power of concurrence has been introduced in \cite{jll}: for $\alpha\geq 2$ and any multiqubit state $\rho_{A|B_0B_1\cdots B_{N-1}}$,
\begin{eqnarray}\label{pfth24}
C^\alpha_{A|B_0B_1\cdots B_{N-1}}\geq\sum_{j=0}^{N-1} (2^{\frac{\alpha}{2}}-1)^jC^\alpha_{AB_j},
\end{eqnarray}
if $C_{AB_i}\geq C_{A|B_{i+1}\cdots B_{N-1}}$ for $i=0,1,\cdots,N-2$. Though inequality (\ref{th21}) is equivalent to inequality (\ref{pfth24}) for any multiqubit states, but the constraint (\ref{th22}) in Theorem 2 is less strict than $C_{AB_i}\geq C_{A|B_{i+1}\cdots B_{N-1}}$ for (\ref{pfth24}), which is to show that Theorem 2 applies to more general states than inequality (\ref{pfth24}). Moreover,  inequality (\ref{th21}) is obviously better, compared with the result (\ref{mo2}) in \cite{JZX}.

\section{tight polygamy constraints of multiqubit entanglement}

As a dual property to the inequality (\ref{th21}) of Theorem 2, we now provide a class of polygamy inequalities of multiqubit entanglement in terms of concurrence and concurrence of assistance, and the Hamming weight of the binary vectors associated with the distribution of subsystems. We first give two lemmas.

{[\bf Lemma 3]}. For any real numbers $x$ and $t$, $0\leq t \leq 1$, $0\leq x \leq 1$, we have $(1+t)^x\leq 1+(2^{x}-1)t^x$.

{\sf [Proof].} Let $f(x,y)=(1+y)^x-y^x$ with $0\leq x\leq 1,~y\geq 1$. Then $\frac{\partial f}{\partial y}=x[(1+y)^{x-1}-y^{x-1}]\leq 0$. Therefore, $f(x,y)$ is an decreasing function of $y$, i.e., $f(x,y)\leq f(x,1)=2^x-1$. Set $y=\frac{1}{t},~0<t\leq 1$, we obtain $(1+t)^x\leq 1+(2^x-1)t^x$. When $t=0$, the inequality is trivial. $\Box$

{[\bf Lemma 4]}. For any $2\otimes2\otimes2^{n-2}$ mixed state $\rho\in \mathds{H}_A\otimes \mathds{H}_{B}\otimes \mathds{H}_{C}$, if $C_{AB}\geq C_{AC}$, we have
\begin{equation}\label{le4}
  C^\beta_{A|BC}\leq  {C_a^\beta}_{AB}+(2^{\frac{\beta}{2}}-1){C_a^\beta}_{AC},
\end{equation}
for all $0\leq\beta\leq2$.

{[\sf Proof]}. It has been shown that $C^2_{A|BC}\leq {C_a^2}_{AB}+{C_a^2}_{AC}$ for arbitrary $2\otimes2\otimes2^{n-2}$ tripartite state $\rho_{ABC}$ \cite{GSB}. Then, if ${C_a}_{AB}\geq {C_a}_{AC}$, we have
\begin{eqnarray*}
  C^\beta_{A|BC}
  &&\leq ({C^2_a}_{AB}+{C^2_a}_{AC})^{\frac{\beta}{2}}\\
  &&={C_a^\beta}_{AB}\left(1+\frac{{C_a^2}_{AC}}{{C^2_a}_{AB}}\right)^{\frac{\beta}{2}} \\
  && \leq {C_a^\beta}_{AB}\left[1+(2^{\frac{\beta}{2}}-1)\left(\frac{{C^2_a}_{AC}}{{C^2_a}_{AB}}\right)^{\frac{\beta}{2}}\right]\\
  &&={C^\beta_a}_{AB}+(2^{\frac{\beta}{2}}-1){C^\beta_a}_{AC},
\end{eqnarray*}
where the second inequality is due to Lemma 3. Here without loss of generality, we have assumed that ${C_a}_{AB}\geq {C_a}_{AC}$ . Moreover, if ${C_a}_{AB}=0$, we have ${C_a}_{AB}={C_a}_{AC}=0$. $\Box$

{[\bf Theorem 3]}. For any $N+1$-qubit state $\rho_{AB_0\cdots B_{N-1}}$ satisfying
\begin{eqnarray}\label{th41}
{C_a}_{AB_j}\geq {C_a}_{AB_{j+1}}\geq 0,
\end{eqnarray}
$j=0,1,\cdots N-2$, we have for $0\leq\beta\leq2$
\begin{eqnarray}\label{th42}
C^\beta_{A|B_0B_1\cdots B_{N-1}}\leq \sum_{j=0}^{N-1} (2^{\frac{\beta}{2}}-1)^{w_H(\vec{j})}{C^\beta_a}_{AB_j}.
\end{eqnarray}

{\sf [Proof].}
The monotonicity of the function $f(x)=x^t$ for $0\leq t\leq 1$ and the generalized monogamy relation based on the concurrence of assistance in inequality (\ref{DCA}) imply that
\begin{eqnarray}\label{pfth41}
C^\beta_{A|B_0B_1\cdots B_{N-1}}\leq\left(\sum_{j=0}^{N-1} {C^2_a}_{AB_j}\right)^\frac{\beta}{2}.
\end{eqnarray}
Therefore, it is sufficient to show that
\begin{eqnarray}\label{pfth42}
\left(\sum_{j=0}^{N-1} {C^2_a}_{AB_j}\right)^\frac{\beta}{2}\leq \sum_{j=0}^{N-1} (2^{\frac{\beta}{2}}-1)^{w_H(\vec{j})}{C^\beta_a}_{AB_j}.
\end{eqnarray}
Similar to the case of Theorem 1, we first prove the inequality (\ref{pfth42}) for the case that $N = 2^n$ by induction. For $n = 1$ and a three-qubit state $\rho_{AB_0B_1}$, we have from Lemma 4,
\begin{equation*}\label{}
  ({C^2_a}_{AB_0}+{C^2_a}_{AB_1})^\frac{\beta}{2}\leq  {C^\beta_a}_{AB_0}+(2^{\frac{\beta}{2}}-1){C^\beta_a}_{AB_1},
\end{equation*}
which gives rise to inequality (\ref{pfth42}) for $n=1$.

Now we assume that the inequality (\ref{pfth42}) is true for $N=2^{n-1}$, $n \geq 2$. For an $(N+1)$-qubit state $\rho_{AB_0\cdots B_{N-1}}$, we have
\begin{eqnarray}\label{pfth43}
&&\left(\sum_{j=0}^{N-1} {C^2_a}_{AB_j}\right)^\frac{\beta}{2}\nonumber\\
&&= \left(\sum_{j=0}^{2^{n-1}-1} {C^2_a}_{AB_j}\right)^\frac{\beta}{2}\left(1+\frac{\sum_{j=2^{n-1}}^{2^n-1} {C^2_a}_{AB_j}}{\sum_{j=0}^{2^{n-1}-1} {C^2_a}_{AB_j}}\right)^\frac{\beta}{2}.
\end{eqnarray}

From (\ref{th41}) we have
\begin{eqnarray}\label{pfth44}
\sum_{j=2^{n-1}}^{2^n-1} {C^2_a}_{AB_j}\leq\sum_{j=0}^{2^{n-1}-1} {C^2_a}_{AB_j}.
\end{eqnarray}
Using Lemma 3 we get
\begin{eqnarray}\label{pfth45}
\left(\sum_{j=0}^{N-1} {C^2_a}_{AB_j}\right)^\frac{\beta}{2} &&\leq\left(\sum_{j=0}^{2^{n-1}-1} {C^2_a}_{AB_j}\right)^\frac{\beta}{2}\nonumber\\
&&+(2^{\frac{\beta}{2}}-1)\left(\sum_{j=2^{n-1}}^{2^n-1} {C^2_a}_{AB_j}\right)^\frac{\beta}{2}.
\end{eqnarray}
Because each of the summations on the right-hand side of inequality (\ref{pfth45}) is a summation of $2^{n-1}$ terms, the induction hypothesis assures that
\begin{eqnarray}\label{pfth46}
\left(\sum_{j=0}^{2^{n-1}-1} {C^2_a}_{AB_j}\right)^\frac{\beta}{2}\leq \sum_{j=0}^{2^{n-1}-1} (2^{\frac{\beta}{2}}-1)^{w_H(\vec{j})}{C^\beta_a}_{AB_j},
\end{eqnarray}
and
\begin{eqnarray}\label{pfth47}
\left(\sum_{j=2^{n-1}}^{2^{n}-1} {C^2_a}_{AB_j}\right)^\frac{\beta}{2}\leq \sum_{j=2^{n-1}}^{2^{n}-1} (2^{\frac{\beta}{2}}-1)^{w_H(\vec{j})-1}{C^\beta_a}_{AB_j}.
\end{eqnarray}
From inequality (\ref{pfth45}) together with inequalities (\ref{pfth46}) and (\ref{pfth47}),
we have
\begin{eqnarray}\label{pfth48}
\left(\sum_{j=0}^{2^{n}-1} {C^2_a}_{AB_j}\right)^\frac{\beta}{2}\leq \sum_{j=0}^{2^{n}-1} (2^{\frac{\beta}{2}}-1)^{w_H(\vec{j})}{C^\beta_a}_{AB_j} ,
\end{eqnarray}
which recovers inequality (\ref{pfth42}).

Now let us consider an arbitrary positive integer $N$ and an $(N + 1)$-qubit state $\rho_{AB_0\cdots B_{N-1}}$. Note that we can always assume that $0\leq N\leq 2^n$ for some $n$. Consider the $(2^n+1)$-qubit state $\rho'_{AB_0\cdots B_{2^n-1}}$ in Eq. (\ref{pfth18}). From (\ref{pfth41}) and (\ref{pfth48})
we have
\begin{eqnarray}\label{pfth49}
C^\beta(\rho'_{A|B_0B_1\cdots B_{2^n-1}})\leq \sum_{j=0}^{2^n-1} (2^{\frac{\beta}{2}}-1)^{w_H(\vec{j})}C^\beta_a(\sigma_{AB_j}),
\end{eqnarray}
where $\sigma_{AB_j}$ is the two-qubit reduced density matrix of $\rho'_{AB_0\cdots B_{2^n-1}}$, $j=0,1,\cdots,2^n-1$.

Moreover, $\rho'_{A|B_0B_1\cdots B_{2^n-1}}$ is a product state of $\rho_{AB_0\cdots B_{N-1}}$ and $\delta_{B_N\cdots B_{2^n-1}}$, which implies that
$C_a(\rho'_{A|B_0B_1\cdots B_{2^n-1}})=C_a(\rho_{A|B_0B_1\cdots B_{N-1}})$,
$C_a(\sigma_{AB_j})=0$ for $j=N,\cdots,2^n-1$, and
$\sigma_{AB_j}=\rho_{AB_j}$ for $j=0,1,\cdots,N-1$.
Therefore, from inequality (\ref{pfth49}) we have
\begin{eqnarray}\label{pfth413}
&&C^\beta(\rho_{A|B_0B_1\cdots B_{N-1}})\nonumber\\
&&=C^\beta(\rho'_{A|B_0B_1\cdots B_{2^n-1}})\nonumber\\
&&\leq \sum_{j=0}^{2^n-1} (2^{\frac{\beta}{2}}-1)^{w_H(\vec{j})}{C^\beta_a}(\sigma_{AB_j})\nonumber\\
&&=\sum_{j=0}^{N-1} (2^{\frac{\beta}{2}}-1)^{w_H(\vec{j})}{C^\beta_a}(\rho_{AB_j}),
\end{eqnarray}
which completes the proof. $\Box$

Theorem 3 gives a new class of polygamy relations for multiqubit states, which includes (\ref{DCA}) as a special case: (\ref{th42}) reduces to (\ref{DCA}) when $\beta=2$.
Similar to the improvement from the inequality (\ref{th12}) to the inequality (\ref{th21}), we can analogously improve the polygamy inequality of Theorem 3 to a tighter inequality under certain condition on
the two-qubit entanglement of assistance.

{\bf [Theorem 4]}. For any multiqubit state $\rho_{AB_0\cdots B_{N-1}}$
conditioned that
\begin{eqnarray}\label{th52}
{C^2_a}_{AB_i}\geq\sum_{j=i+1}^{N-1} {C^2_a}_{AB_j}
\end{eqnarray}
for $i=0,1,\cdots N-2$, we have
\begin{eqnarray}\label{th51}
C^\beta_{A|B_0B_1\cdots B_{N-1}}\leq\sum_{j=0}^{N-1} (2^{\frac{\beta}{2}}-1)^j{C^\beta_a}_{AB_j}
\end{eqnarray}
for $0\leq\beta\leq 2$.

{\sf [Proof].}
From the inequality (\ref{le4}), we have
\begin{eqnarray*}\label{pfth51}
&&C^{\beta}_{A|B_0B_1\cdots B_{N-1}}\nonumber\\
&&\leq  {C^{\beta}_a}_{AB_0}+(2^{\frac{\beta}{2}}-1) \left(\sum_{j=1}^{N-1}{C^2_a}_{AB_j}\right)^\frac{\beta}{2}\nonumber\\
&&\leq {C^{\beta}_a}_{AB_0}+(2^{\frac{\beta}{2}}-1){C^{\beta}_a}_{AB_1}
 +(2^{\frac{\beta}{2}}-1)^2 \left(\sum_{j=2}^{N-1}{C^2_a}_{AB_j}\right)^\frac{\beta}{2}\nonumber\\
&& \leq \cdots\nonumber\\
&&\leq {C^{\beta}_a}_{AB_0}+(2^{\frac{\beta}{2}}-1){C^{\beta}_a}_{AB_1}+\cdots+(2^{\frac{\beta}{2}}-1)^{N-1}{C^{\beta}_a}_{AB_{N-1}}.
\end{eqnarray*}
$\Box$

From inequality (\ref{pfth22}), $w_H(\vec{j})\leq j$, we have
\begin{eqnarray}\label{pfth23}
&&C^\beta_{A|B_0B_1\cdots B_{N-1}}\nonumber\\
&&\leq\sum_{j=0}^{N-1} (2^{\frac{\beta}{2}}-1)^j{C^\beta_a}_{AB_j}\nonumber\\
&&\leq\sum_{j=0}^{N-1} (2^{\frac{\beta}{2}}-1)^{w_H(\vec{j})}{C^\beta_a}_{AB_j},
\end{eqnarray}
for any $0\leq\beta\leq 2$. Thus, inequality (\ref{th51}) of Theorem 4 is tighter than the inequality (\ref{th41}) of Theorem 3 for any $0\leq\beta\leq 2$, and any multiqubit state $\rho_{AB_0B_1\cdots B_{N-1}}$ satisfying the conditions (\ref{th52}).

Generally, the conditions (\ref{th52}) is not always satisfied. In general, we have the following
conclusion.

{[\bf Theorem 5]}. For any multiqubit state $\rho_{AB_0\cdots B_{N-1}}$, if
${C_a^2}_{AB_i}\geq \sum_{k=i+1}^{N-1}{C_a^2}_{AB_k}$ for $i=0, 1, \cdots, m$, and
${C_a^2}_{AB_j}\leq \sum_{k=j+1}^{N-1}{C_a^2}_{AB_k}$ for $j=m+1,\cdots,N-2$,
$\forall$ $1\leq m\leq N-3$, $N\geq 4$, we have
\begin{eqnarray}\label{thn1}
&&C^\beta_{A|B_0B_1\cdots B_{N-1}}\leq \nonumber \\
&&{C_a^\beta}_{AB_0}+(2^{\frac{\beta}{2}}-1) {C_a^\beta}_{AB_1}+\cdots+(2^{\frac{\beta}{2}}-1)^{m}{C_a^\beta}_{AB_m}\nonumber\\
&&+(2^{\frac{\beta}{2}}-1)^{m+2}({C_a^\beta}_{AB_{m+1}}+\cdots+{C_a^\beta}_{AB_{N-2}}) \nonumber\\
&&+(2^{\frac{\beta}{2}}-1)^{m+1}{C_a^\beta}_{AB_{N-1}},
\end{eqnarray}
for all $0\leq\beta\leq2$.

{\sf [Proof].} From (\ref{DCA}) and (\ref{le4}), we have
\begin{eqnarray}\label{pfn1}
&&C^{\beta}_{A|B_1B_2\cdots B_{N-1}}\nonumber\\
&&\leq {C_a^\beta}_{AB_0}+(2^{\frac{\beta}{2}}-1) \left(\sum_{i=1}^{N-1}{C^2_a}_{AB_i}\right)^\frac{\beta}{2}\nonumber\\
&&\leq {C_a^\beta}_{AB_0}+(2^{\frac{\beta}{2}}-1){C_a^\beta}_{AB_1}
 +(2^{\frac{\beta}{2}}-1)^2\left(\sum_{i=2}^{N-1}{C^2_a}_{AB_i}\right)^\frac{\beta}{2}\nonumber\\
&& \leq \cdots\nonumber\\
&&\leq {C_a^\beta}_{AB_0}+(2^{\frac{\beta}{2}}-1){C_a^\beta}_{AB_1}+\cdots+(2^{\frac{\beta}{2}}-1)^{m}{C_a^\beta}_{AB_m}\nonumber\\
&&~~~~+(2^{\frac{\beta}{2}}-1)^{m+1} \left(\sum_{i={m+1}}^{N-1}{C^2_a}_{AB_i}\right)^\frac{\beta}{2}.
\end{eqnarray}
Similarly, as ${C_a^2}_{AB_j}\leq \sum_{k=j+1}^{N-1}{C_a^2}_{AB_k}$ for $j=m+1,\cdots,N-2$, we get
\begin{eqnarray}\label{pfn2}
&&\left(\sum_{i={m+1}}^{N-1}{C^2_a}_{AB_i}\right)^\frac{\beta}{2} \nonumber\\
&&\leq (2^{\frac{\beta}{2}}-1){C_a^\beta}_{AB_{m+1}}+\left(\sum_{i={m+2}}^{N-1}{C^2_a}_{AB_i}\right)^\frac{\beta}{2}\nonumber\\
&&\leq (2^{\frac{\beta}{2}}-1)({C_a^\beta}_{AB_{m+1}}+\cdots+{C_a^\beta}_{AB_{N-2}})\nonumber\\
&&~~~~+{C_a^\beta}_{AB_{N-1}}.
\end{eqnarray}
Combining (\ref{pfn1}) and (\ref{pfn2}), we have Theorem 5. $\Box$

Theorem 5 gives another polygamy relation based on the concurrence of assistance. Compared the inequality (\ref{th51}) of Theorem 4 with (\ref{thn1}) of Theorem 5, (\ref{th51}) is better than (\ref{thn1}), obviously. But, for some classes of states, Theorem 5 is better than Theorem 3, for that those classes of states do not satisfy the conditions (\ref{th52}) in Theorem 4.

{\it Example 1}. Let us consider the 4-qubit generlized $W$-class states,
\begin{eqnarray}\label{W4}
|W\rangle_{AB_1B_2B_3}=\frac{1}{2}(|1000\rangle+|0100\rangle+|0010\rangle+|0001\rangle).
\end{eqnarray}
We have $C(|W\rangle_{A|B_1B_2B_3})=\frac{\sqrt{3}}{2}$, ${C_a}_{AB_i}=\frac{1}{2}$, $i=1,2,3$. It is easy to see ${C_a}_{AB_i}$ do not satisfy the conditions (\ref{th52}) in Theorem 4. From inequality (\ref{th42}) of Theorem 3, we have $C^\beta(|W\rangle_{A|B_1B_2B_3})\leq ((2^{\frac{\beta}{2}}-1)^2+2)(\frac{1}{2})^\beta$. From inequality (\ref{thn1}) of Theorem 5, we have $C^\beta(|W\rangle_{A|B_1B_2B_3})\leq (2^{\frac{\beta}{2}+1}-1)(\frac{1}{2})^\beta$.
One can see that the inequality (\ref{thn1}) is better than (\ref{th42}) for $0\leq \beta\leq2$; see Fig. 1.

\begin{figure}
  \centering
  \includegraphics[width=7cm]{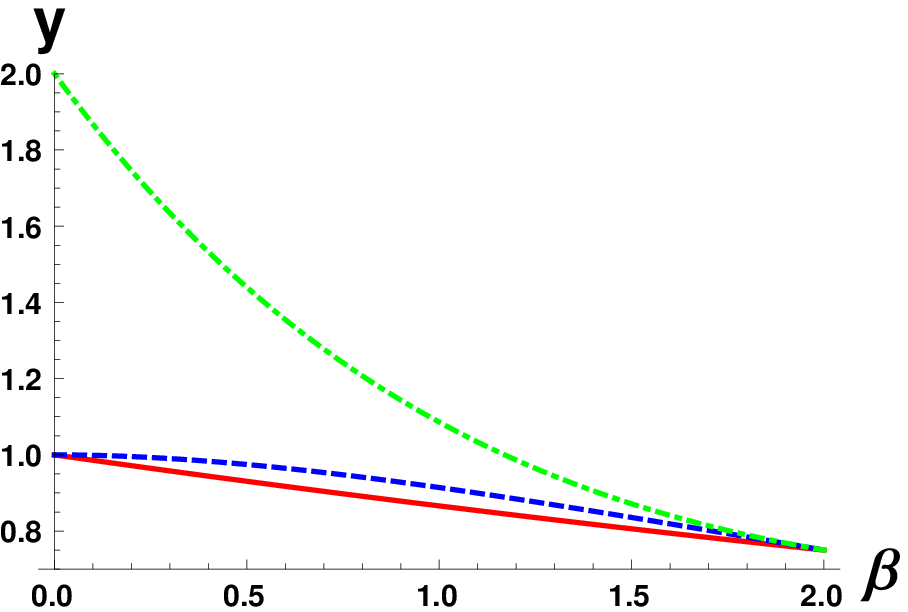}\\
  \caption{$y$ is the value of $C(|W\rangle_{AB_1B_2B_3})$. Solid (red) line is the exact value of $C(|W\rangle_{AB_1B_2B_3})$, dashed (blue) line is the upper bound of $C(|W\rangle_{AB_1B_2B_3})$ in (\ref{thn1}), dot-dashed (green) line is the upper bound in (\ref{th42}) for $0\leq \beta\leq2$.}\label{1}
\end{figure}

\section{MONOGAMY and polygamy RELATIONS for GENERAL quantum correlations}

We have studied the monogamy and polygamy properties related to concurrence and concurrence of assistance.
Now we consider general measures of quantum correlations.
Let $\mathcal{Q}$ be an arbitrary measure of quantum correlation for bipartite systems. $\mathcal{Q}$ is said to be monogamous if it satisfies the following inequality for an $N$-partite quantum state $\rho_{AB_1B_2,\cdots,B_{N-1}}$ \cite{ARA},
\begin{eqnarray}\label{q}
&&\mathcal{Q}(\rho_{A|B_1B_2,\cdots,B_{N-1}})\nonumber\\
&&\geq\mathcal{Q}(\rho_{AB_1})+\mathcal{Q}(\rho_{AB_2})+\cdots+\mathcal{Q}(\rho_{AB_{N-1}}),
\end{eqnarray}
where $\rho_{AB_i}$, $i=1,...,N-1$, are the reduced density matrices, $\mathcal{Q}(\rho_{A|B_1B_2,\cdots,B_{N-1}})$ denotes the quantum correlation $\mathcal{Q}$ of the state $\rho_{AB_1B_2,\cdots,B_{N-1}}$ under bipartite partition  $A|B_1B_2,\cdots,B_{N-1}$. For simplicity, we denote $\mathcal{Q}(\rho_{AB_i})$ by $\mathcal{Q}_{AB_i}$, and $\mathcal{Q}(\rho_{A|B_1B_2,\cdots,B_{N-1}})$ by $\mathcal{Q}_{A|B_1B_2,\cdots,B_{N-1}}$.
One can define the $\mathcal{Q}$-monogamy score for the $N$-partite state $\rho_{AB_1B_2,\cdots,B_{N-1}}$,
\begin{eqnarray}\label{}
\delta_{\mathcal{Q}}=\mathcal{Q}_{A|B_1B_2,\cdots,B_{N-1}}-\sum_{i=1}^{N-1}\mathcal{Q}_{AB_i}.
\end{eqnarray}
Non-negativity of $\delta_{\mathcal{Q}}$ for all quantum states implies the monogamy of $\mathcal{Q}$. For instance, the square of the concurrence has been shown to be monogamous \cite{AKE,SSS} for all multi-qubit states.
However, there are other measures like entanglement of formation, quantum discord, and quantum work deficit which are known to be nonmonogamous for pure three-qubit states \cite{GLGP, RPAK}.

Given any quantum correlation measure that is non-monogamic for a multipartite quantum state, it is always possible to find a monotonically increasing function of the measure which is monogamous for the same state \cite{SPAU}.
It has been proved that for arbitrary dimensional tripartite states, there exists $\gamma\in R~(\gamma\geq1)$ such that a quantum correlation measure $\mathcal{Q}$ satisfies the following monogamy relation \cite{SPAU}
\begin{eqnarray}\label{aq}
\mathcal{Q}^\gamma_{A|BC}\geq\mathcal{Q}^\gamma_{AB}+\mathcal{Q}^\gamma_{AC}.
\end{eqnarray}

In the following, we denote $\gamma$ the minimal value such that $\mathcal{Q}$ satisfies the above inequality.  Using the inequality $(1+t)^x\geq 1+t^x$ for $x\geq1,~0\leq t\leq1$, it is easy to generalize the result (\ref{aq}) to the $N$ partite case,
\begin{eqnarray}\label{mq}
\mathcal{Q}^\gamma_{A|B_0B_1,\cdots,B_{N-1}}\geq \sum_{i=0}^{N-1}\mathcal{Q}_{AB_i}^\gamma,
\end{eqnarray} where $i=0,1,\cdots, N-1$. Using the similar method of Theorem 1, and combining inequality (\ref{mq}) with Lemma 1, we have the following result.

{[\bf Theorem 6]}. For any $N+1$-qubit state $\rho_{AB_0\cdots B_{N-1}}$ satisfying
\begin{eqnarray}\label{th71}
\mathcal{Q}_{AB_j}\geq \mathcal{Q}_{AB_{j+1}}\geq 0,
\end{eqnarray}
$j=0,1,\cdots N-2$, we have for $\alpha\geq\gamma$
\begin{eqnarray}\label{th72}
\mathcal{Q}^\alpha_{A|B_0B_1\cdots B_{N-1}}\geq \sum_{j=0}^{N-1} (2^{\frac{\alpha}{\gamma}}-1)^{w_H(\vec{j})}\mathcal{Q}^\alpha_{AB_j}.
\end{eqnarray}

Using the similar method of Theorem 2, and combining inequality (\ref{mq}) with Lemma 2, we have the following results.

{\bf [Theorem 7]}. For any multiqubit state $\rho_{AB_0\cdots B_{N-1}}$, we have
\begin{eqnarray}\label{th81}
\mathcal{Q}^\alpha_{A|B_0B_1\cdots B_{N-1}}\geq\sum_{j=0}^{N-1} (2^{\frac{\alpha}{\gamma}}-1)^j\mathcal{Q}^\alpha_{AB_j},
\end{eqnarray}
if
\begin{eqnarray}\label{th82}
\mathcal{Q}^\gamma_{AB_i}\geq\sum_{j=i+1}^{N-1} \mathcal{Q}^\gamma_{AB_j},
\end{eqnarray}
for $i=0,1,\cdots N-2$, $\alpha\geq \gamma$.

We provide a class of general polygamy inequalities in terms of powered quantum correlation measure $\mathcal{Q}$ and the Hamming weight of the binary vector related with the distribution of subsystems. Using the similar method of Theorem 3, and combining the inequality (\ref{mq}) with Lemma 3, we have the following result.

{[\bf Theorem 8]}. For a $N+1$-qubit state $\rho_{AB_0\cdots B_{N-1}}$ satisfying
\begin{eqnarray}\label{th91}
\mathcal{Q}_{AB_j}\geq \mathcal{Q}_{AB_{j+1}}\geq 0,
\end{eqnarray}
$j=0,1,\cdots N-2$, we have for $0\leq\beta\leq \gamma$
\begin{eqnarray}\label{th92}
\mathcal{Q}^\beta_{A|B_0B_1\cdots B_{N-1}}\leq \sum_{j=0}^{N-1} (2^{\frac{\beta}{\gamma}}-1)^{w_H(\vec{j})}{\mathcal{Q}^\beta}_{AB_j}.
\end{eqnarray}

Moreover, using the similar method of Theorem 4, and combining the inequality (\ref{mq}) with Lemma 4, we have the following result.

{\bf [Theorem 9]}. For any multiqubit state $\rho_{AB_0\cdots B_{N-1}}$, we have for $0\leq\beta\leq \gamma$
\begin{eqnarray}\label{th101}
\mathcal{Q}^\beta_{A|B_0B_1\cdots B_{N-1}}\leq\sum_{j=0}^{N-1} (2^{\frac{\beta}{\gamma}}-1)^j{\mathcal{Q}^\beta}_{AB_j}
\end{eqnarray}
if
\begin{eqnarray}\label{th102}
\mathcal{Q}^\gamma_{AB_i}\geq\sum_{j=i+1}^{N-1} \mathcal{Q}^\gamma_{AB_j}
\end{eqnarray}
for $i=0,1,\cdots N-2$.

Similarly, corresponding to Theorem 5, from (\ref{mq}) and Lemma 4 we have the following result.

{[\bf Theorem 10]}. For any multiqubit state $\rho_{AB_0\cdots B_{N-1}}$, if
${\mathcal{Q}^\gamma_{AB_i}}\geq \sum_{k=i+1}^{N-1}{\mathcal{Q}^\gamma}_{A|B_{i+1}}$ for $i=0, 1, \cdots, m$, and
${\mathcal{Q}^\gamma_{AB_j}}\leq \sum_{k=j+1}^{N-1}{\mathcal{Q}^\gamma}_{A|B_{i+1}}$ for $j=m+1,\cdots,N-2$,
$\forall$ $1\leq m\leq N-3$, $N\geq 4$, we have
\begin{eqnarray}\label{th121}
&&\mathcal{Q}^\beta_{A|B_0B_1\cdots B_{N-1}}\leq \nonumber \\
&&\mathcal{Q}^\beta_{AB_0}+(2^{\frac{\beta}{\gamma}}-1) \mathcal{Q}^\beta_{AB_1}+\cdots+(2^{\frac{\beta}{\gamma}}-1)^{m}\mathcal{Q}^\beta_{AB_m}\nonumber\\
&&+(2^{\frac{\beta}{\gamma}}-1)^{m+2}(\mathcal{Q}^\beta_{AB_{m+1}}+\cdots+\mathcal{Q}^\beta_{AB_{N-2}}) \nonumber\\
&&+(2^{\frac{\beta}{\gamma}}-1)^{m+1}\mathcal{Q}^\beta_{AB_{N-1}},
\end{eqnarray}
for all $0\leq\beta\leq\gamma$.

{\sf [Remark 2].} We have presented a universal form of monogamy and polygamy relations for any quantum correlations. Our general monogamy and polygamy relations can be used to any quantum correlation measures like SCREN, entanglement of formation and Tsallis-$q$ entanglement, and
give rise to either tighter monogamy relations than the existing ones for some classes of quantum states, or less restricted conditions on states than the ones for the existing monogamy relations, namely, these monogamy and polygamy relations apply to more general quantum states.

In the following, we take SCREN as an example to show the advantage of our conclusions.

Given a bipartite state $\rho_{AB}$ in $H_A\otimes H_B$, the negativity is defined by \cite{GRF},
$N(\rho_{AB})=(||\rho_{AB}^{T_A}||-1)/2$,
where $\rho_{AB}^{T_A}$ is the partially transposed $\rho_{AB}$ with respect to the subsystem $A$, $||X||$ denotes the trace norm of $X$, i.e $||X||=\mathrm{Tr}\sqrt{XX^\dag}$.
 For the purpose of discussion, we use the following definition of negativity, $ N(\rho_{AB})=||\rho_{AB}^{T_A}||-1$.
For any bipartite pure state $|\psi\rangle_{AB}$, the negativity $ N(\rho_{AB})$ is given by
$N(|\psi\rangle_{AB})=2\sum_{i<j}\sqrt{\lambda_i\lambda_j}=(\mathrm{Tr}\sqrt{\rho_A})^2-1$,
where $\lambda_i$ are the eigenvalues for the reduced density matrix $\rho_A$ of $|\psi\rangle_{AB}$. For a mixed state $\rho_{AB}$, the SCREN is defined by
\begin{equation}\label{nc}
 N_{sc}(\rho_{AB})=[\mathrm{min}\sum_ip_iN(|\psi_i\rangle_{AB})]^2,
\end{equation}
where the minimum is taken over all possible pure state decompositions $\{p_i,~|\psi_i\rangle_{AB}\}$ of $\rho_{AB}$. Similar to the duality between concurrence and concurrence of assistance, we also define a dual quantity to SCREN as
\begin{equation}\label{na}
 N_{sc}^a(\rho_{AB})=[\mathrm{max}\sum_ip_iN(|\psi_i\rangle_{AB})]^2,
\end{equation}
which we refer to as the SCREN of assistance (SCRENoA), and the maximum is taken over all possible pure state decompositions $\{p_i,~|\psi_i\rangle_{AB}\}$ of $\rho_{AB}$. For convenience, we denote ${N_{sc}}_{AB_i}=N_{sc}(\rho_{AB_i})$ the SCREN of $\rho_{AB_i}$ and ${N_{sc}}_{AB_0,B_1\cdots,B_{N-1}}=N_{sc}(|\psi\rangle_{AB_0\cdots B_{N-1}})$.

In \cite{j012334} it has been shown that
\begin{eqnarray}\label{n1}
{N_{sc}}_{A|B_0B_1\cdots B_{N-1}}\geq \sum_{j=0}^{N-1}{N_{sc}}_{AB_j},
\end{eqnarray}
and
\begin{eqnarray}\label{n2}
{N_{sc}^a}_{A|B_0B_1\cdots B_{N-1}}\leq \sum_{j=0}^{N-1}{N_{sc}^a}_{AB_j}.
\end{eqnarray}

It is further improved that for $\alpha\geq 1$ \cite{j012334},
\begin{eqnarray}\label{nn1}
{N^\alpha_{sc}}_{A|B_0B_1\cdots B_{N-1}}\geq \sum_{j=0}^{N-1}\alpha^{w_H(\vec{j})}{N^\alpha_{sc}}_{AB_j},
\end{eqnarray}
and for  $0\leq\beta\leq 1$,
\begin{eqnarray}\label{nn2}
({N_{sc}^a}_{A|B_0B_1\cdots B_{N-1}})^\beta\leq \sum_{j=0}^{N-1} \beta^{w_H(\vec{j})}({N_{sc}^a}_{AB_j})^\beta,
\end{eqnarray}
\begin{eqnarray}\label{nn3}
({N_{sc}^a}_{A|B_0B_1\cdots B_{N-1}})^\beta\leq \sum_{j=0}^{N-1} \beta^j({N_{sc}^a}_{AB_j})^\beta.
\end{eqnarray}

Combining inequality  (\ref{th72}) and (\ref{n1}), we obtain a tighter monogamy relation of SCREN $(\gamma=1)$,
\begin{eqnarray}\label{n3}
{N^\alpha_{sc}}_{A|B_0B_1\cdots B_{N-1}}\geq \sum_{j=0}^{N-1}(2^\alpha-1)^{w_H(\vec{j})}{N^\alpha_{sc}}_{AB_j},
\end{eqnarray}
which is better than the result (\ref{nn1}) in \cite{j012334}, since ${N^\alpha_{sc}}_{A|B_0B_1\cdots B_{N-1}}\geq \sum_{j=0}^{N-1}\alpha^{w_H(\vec{j})}{N^\alpha_{sc}}_{AB_j}$, as $(2^\alpha-1)^{w_H(\vec{j})}\geq\alpha^{w_H(\vec{j})}$ for $\alpha\geq 1$.

{\it Example 2}. Let us consider the $3\otimes 2\otimes 2$ quantum state \cite{j012329},
\begin{eqnarray}\label{ex2}
|\psi\rangle_{ABC}=\frac{1}{\sqrt{6}}(\sqrt{2}|100\rangle+\sqrt{2}|101\rangle+|200\rangle+|211\rangle),
\end{eqnarray}
which violates the tangle-based monogamy inequality. But the inequality (\ref{n3}) still holds.
We have ${N_{sc}}_{ABC}=4$ and ${N_{sc}}_{AB}={N_{sc}}_{AC}=\frac{8}{9}$. Therefore, the SCERN-based monogamy inequality (\ref{n3}) is given by ${N^\alpha_{sc}}_{ABC}\geq {N^\alpha_{sc}}_{AB}+(2^\alpha-1)^{w_H(\vec{j})}{N^\alpha_{sc}}_{AC}= \left(\frac{8}{9}\right)^\alpha2^\alpha$, while the result (\ref{nn1}) in \cite{j012334} is given by ${N^\alpha_{sc}}_{ABC}\geq{N^\alpha_{sc}}_{AB}+(\alpha)^{w_H(\vec{j})}{N^\alpha_{sc}}_{AC}= (1+\alpha) \left(\frac{8}{9}\right)^\alpha$. One can see that our result is better than that in  \cite{j012334} for $\alpha\geq 1$; see Fig. 2.

\begin{figure}
  \centering
  \includegraphics[width=7cm]{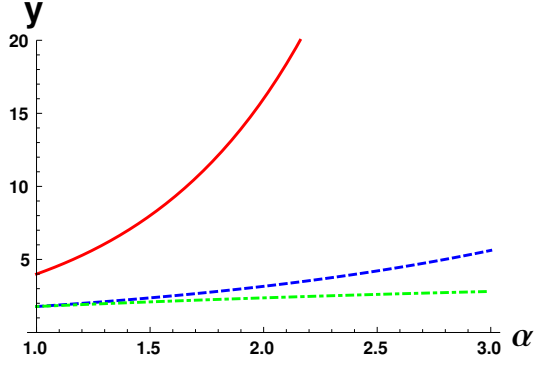}\\
  \caption{$y$ is the value of $N_{sc}(|\psi\rangle_{ABC})$. Solid (red) line is the exact value of $N_{sc}(|\psi\rangle_{ABC})$, dashed (blue) line is the lower bound of $N_{sc}(|\psi\rangle_{ABC})$ in (\ref{n3}), dot-dashed (green) line is the lower bound in \cite{j012334} for $\alpha\geq1$.}\label{1}
\end{figure}

Similarly, combining the inequalities (\ref{th92}) and (\ref{n2}), we obtain a tighter polygamy relation of SCRENoA $(\gamma=1)$,
\begin{eqnarray}\label{n4}
({N_{sc}^a}_{A|B_0B_1\cdots B_{N-1}})^\beta\leq \sum_{j=0}^{N-1} (2^\beta-1)^{w_H(\vec{j})}({N_{sc}^a}_{AB_j})^\beta,
\end{eqnarray}
which is better than the result (\ref{nn2}) in \cite{j012334}, because $({N_{sc}^a}_{A|B_0B_1\cdots B_{N-1}})^\beta\leq \sum_{j=0}^{N-1} \beta^{w_H(\vec{j})}({{N_{sc}^a}_{AB_j})^\beta}$, as $(2^\beta-1)^{w_H(\vec{j})}\leq\beta^{w_H(\vec{j})}$ for $0\leq\beta\leq 1$.

{\it Example 3}. Let us consider the 4-qubit generlized $W$-class states (\ref{W4}).
We have ${N_{sc}^a}_{A|B_1B_2B_3}=\frac{3}{4}$, ${N_{sc}^a}_{AB_i}=\frac{1}{4}$, $i=1,2,3$. From our result (\ref{n4}) we have $({N_{sc}^a}_{A|B_1B_2B_3})^\beta\leq (2^\beta+1)(\frac{1}{4})^\beta$. From the result (\ref{nn2}) in \cite{j012334}, one has $({N_{sc}^a}_{A|B_1B_2B_3})^\beta\leq (2+\beta)(\frac{1}{4})^\beta$. One can see that our result is better than that in \cite{j012334} for $0\leq \beta\leq1$; see Fig. 3.
\begin{figure}
  \centering
  \includegraphics[width=7cm]{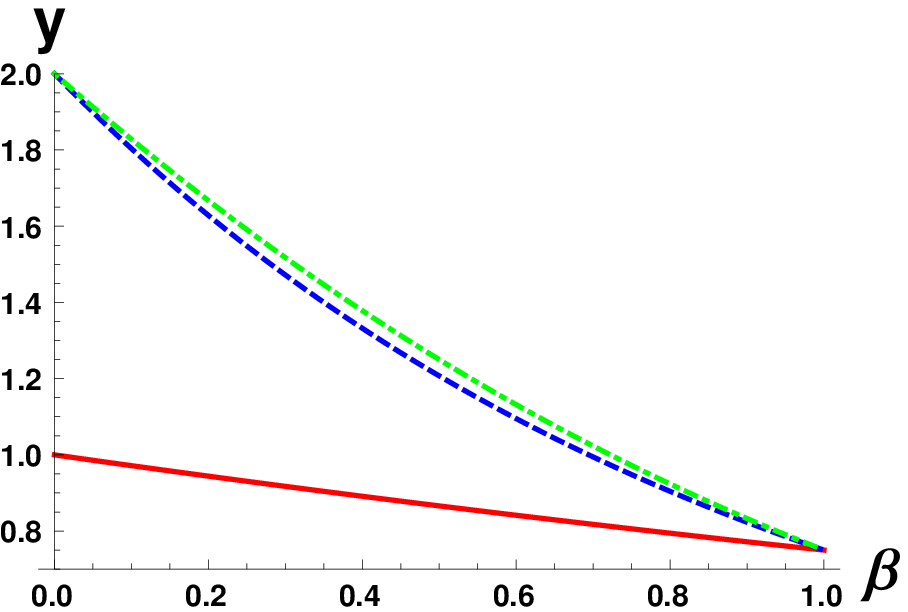}\\
  \caption{$y$ is the value of ${N_{sc}^a}_{A|B_1B_2B_3}$. Solid (red) line is the exact value of ${N_{sc}^a}_{A|B_1B_2B_3}$, dashed (blue) line is the upper bound of ${N_{sc}^a}_{A|B_1B_2B_3}$ in (\ref{n4}), dot-dashed (green) line is the upper bound in \cite{j012334} for $0\leq\beta\leq1$.}\label{1}
\end{figure}

Combining the inequalities (\ref{th101}) and (\ref{n2}), we can also obtain a polygamy relation based on SCRENoA $(\gamma=1)$,
\begin{eqnarray}\label{n5}
({N_{sc}^a}_{A|B_0B_1\cdots B_{N-1}})^\beta\leq\sum_{j=0}^{N-1} (2^\beta-1)^j({N_{sc}^a}_{AB_j})^\beta,
\end{eqnarray}
for all $0\leq\beta\leq1$.
As $2^\beta-1\leq \beta$ for $0\leq\beta\leq1$, the inequality (\ref{n5}) is better than the result (\ref{nn3}) in \cite{j012334}.

Combining inequality (\ref{th121}) and (\ref{n2}), we can get another polygamy relation based on SCRENoA $(\gamma=1)$,
\begin{eqnarray}\label{n6}
&&({N_{sc}^a}_{A|B_0B_1\cdots B_{N-1}})^\beta\leq \nonumber \\
&&({N_{sc}^a}_{AB_0})^\beta+(2^{\frac{\beta}{2}}-1) ({N_{sc}^a}_{AB_1})^\beta+\cdots\nonumber\\
&&+(2^{\frac{\beta}{2}}-1)^{m}(({N_{sc}^a}_{AB_m})^\beta+(2^{\frac{\beta}{2}}-1)^{m+2}(({N_{sc}^a}_{AB_{m+1}})^\beta\nonumber\\
&&+\cdots+({N_{sc}^a}_{AB_{N-2}})^\beta)+(2^{\frac{\beta}{2}}-1)^{m+1}({N_{sc}^a}_{AB_{N-1}})^\beta,
\end{eqnarray}
for all $0\leq\beta\leq1$.

Inequality (\ref{n6}) gives another polygamy relation based on SCRENoA.
(\ref{n5}) is better than (\ref{n6}) obviously for some classes of states.
However, for some other classes of states which do not satisfy the constraint (\ref{n5}), (\ref{n6}) is better than (\ref{n4}).

{\it Example 4}. Let us again consider the 4-qubit generlized $W$-class states (\ref{W4}).
We have ${N_{sc}^a}_{A|B_1B_2B_3}=\frac{3}{4}$, ${N_{sc}^a}_{AB_i}=\frac{1}{4}$, $i=1,2,3$. From inequality (\ref{n4}), we have $({N_{sc}^a}_{A|B_1B_2B_3})^\beta\leq \left[2+(2^\beta-1)^2\right](\frac{1}{4})^\beta$. From inequality (\ref{n6}), we have $({N_{sc}^a}_{A|B_1B_2B_3})^\beta\leq (2^{\beta+1}-1)(\frac{1}{4})^\beta$.
From the result (\ref{nn3}) in \cite{j012334}, one has $({N_{sc}^a}_{A|B_1B_2B_3})^\beta\leq (2+\beta^2)(\frac{1}{4})^\beta$. One can see that our results are better than that in \cite{j012334}, and inequality (\ref{n6}) is better than (\ref{n4}) for $0\leq \beta\leq1$; see Fig. 4.
\begin{figure}
  \centering
  \includegraphics[width=7cm]{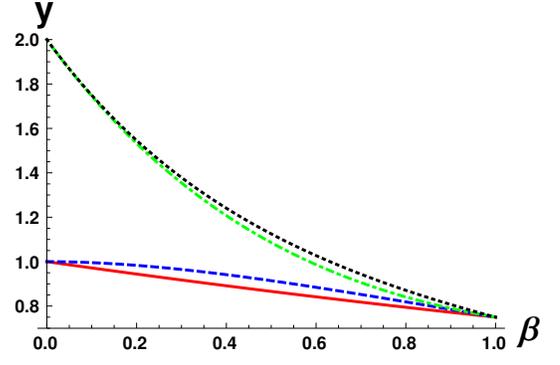}\\
  \caption{$y$ is the value of ${N_{sc}^a}_{A|B_1B_2B_3}$. Solid (red) line is the exact value of ${N_{sc}^a}_{A|B_1B_2B_3}$, dashed (blue) line is the upper bound of ${N_{sc}^a}_{A|B_1B_2B_3}$ in (\ref{n6}), dot-dashed (green) line is the upper bound in (\ref{n4}), dotted (black) line is the upper bound in \cite{j012334} for $0\leq\beta\leq1$.}\label{1}
\end{figure}

\section{conclusion}
Entanglement monogamy is a fundamental property of multipartite entangled states.
We have provided a characterization of multiqubit entanglement constraints in terms of concurrence.
By using the Hamming weight of the binary vectors related to the individual subsystems, we have established
a class of monogamy inequalities of multiqubit entanglement based on the $\alpha$th power of concurrence for $\alpha\geq 2$.
We have also established a class of polygamy inequalities of multiqubit entanglement in terms of
the $\beta$th power of concurrence and concurrence of assistance for $0\leq \beta\leq2$.
Moveover, the monogamy and polygamy inequalities for general quantum correlations have been presented.
Applying these results to the quantum correlations such as SCREN, entanglement of formation, and Tsallis-$q$ entanglement, one obtains tighter monogamy relations
than the existing ones for some classes of quantum states, or  monogamy relations with less restrictions on the quantum states.
Monogamy relations characterize the distributions of quantum correlations in multipartite systems. Tighter monogamy relations imply finer characterizations
of the quantum correlation distributions. Our approach may also be used to study further the monogamy and polygamy inequalities for other high dimensional quantum systems.

\bigskip
\noindent{\bf Acknowledgments}\, \, This work is supported by the NSF of China under Grant No. 11675113, and NSF of Beijing under No. KZ201810028042.

\end{document}